\documentclass[12pt]{iopart}

\usepackage{amssymb}

\begin{document}

\newcommand \be  {\begin{equation}}
\newcommand \bea {\begin{eqnarray} \nonumber }
\newcommand \ee  {\end{equation}}
\newcommand \eea {\end{eqnarray}}

\title[Pre-freezing in logarithmically
correlated potential]{Pre-freezing of multifractal exponents in
Random Energy Models with logarithmically correlated potential}

\vskip 0.2cm
\author{Yan V Fyodorov\footnote{e-mail: yan.fyodorov@nottingham.ac.uk}}

 \noindent\small{ School of Mathematical Sciences,
University of Nottingham, Nottingham NG72RD, England}

\begin{abstract}
Boltzmann-Gibbs measures generated by logarithmically correlated
random potentials are multifractal. We investigate the abrupt
change ("pre-freezing") of multifractality exponents extracted
from the averaged moments of the measure - the so-called inverse
participation ratios. The pre-freezing can be identified with
termination of the disorder-averaged multifractality spectrum.
Naive replica limit employed to study a one-dimensional variant of
the model is shown to break down at the pre-freezing point.
Further insights are possible when employing zero-dimensional and
infinite-dimensional versions of the problem. In particular, the
latter version allows one to identify the pattern of the replica
symmetry breaking responsible for the pre-freezing phenomenon.
\end{abstract}

\maketitle

\section{Introduction}

Investigations of multifractal measures of diverse origin is a
very active field of research in various branches of physics for
several decades \cite{PV}. To set the notations, consider a
certain (e.g. hypercubic) lattice of linear extent $L$ in
$N-$dimensional space, with $M\sim L^N$ standing for the total
number of sites in the lattice. The measures of interest are
usually defined via weights $0\le p_i\le 1$ associated with every
lattice site $i=1,2,\ldots, M$ and normalized as $\sum_{i=1}^M
\,p_i=1$. One can imagine a few different spatial arrangements of
weights $p_i$ accross the lattice sites. In the case of {\it
simply extended} measures the weights are of similar magnitude at
each lattice site, the normalisation condition then implying the
scaling $p_i\sim M^{-1}$ in the large-$M$ limit. As a
generalisation of the above example one can imagine the non-zero
weights $p_i$ supported evenly on a fractal subset of lattice
sites of effective dimension $0\le N_{ef}<N$. In the limiting case
of $N_{ef}=0$ we then deal with {\it localized} measures
characterized by the weights $p_i$ essentially different from zero
only inside one or few blobs of finite total volume. In such a
situation weights stay finite even when $M\to \infty$, that is
$p_i=O(M^0)$. Finally, in the most interesting case of {\it
multifractal} measures the weights scale differently at different
sites: $p_i\sim M^{-\alpha_i}$ \footnote{Usually one defines
exponents via the relation $p_i\sim L^{-N\alpha_i}$ i.e. by the
reference to linear scale $L$ instead of the total number of sites
$M\sim L^N$. We however find it more convenient to get rid of
trivial spatial dimension factor $N$, and concentrate only on {\it
essential} parameter behaviour.}. The full set of exponents $0\le
\alpha_i<\infty$ can be conveniently characterized by the density
$ \rho(\alpha)=\sum_{i=1}^M\,\delta(\alpha-\alpha_i)$ whose
scaling behaviour in the large-$M$ limit is expected to be
nontrivial: $\rho(\alpha)\sim M^{f(\alpha)}$, with the (convex)
function $f(\alpha)$ known in this context as the {\it
multifractality spectrum}. Note that the total number
$m(\alpha)=\int_{0}^{\alpha}\rho(\alpha)\,d\alpha$ of sites of the
lattice characterised by the scaling exponents $\alpha_i<\alpha$
must satisfy the inequality $m(\alpha)\gtrsim M^{f(\alpha)}\ge 1$,
hence $f(\alpha)\ge 0$. The condition $f(\alpha)= 0$ defines
generically the minimal $\alpha_{-}$ and maximal $\alpha_{+}$
threshold values of the exponents which can be observed in a
typical realisation of disorder. Note that the constraint $p_i\le
1$ implies $\alpha_{-}\ge 0$.

An alternative, frequently more practical way of describing
multifractality  is via the set of exponents $\tau_q$
characterising the large-$M$ behaviour of the so-called inverse
participation ratios (IPR's) $P_q$ which are simply the moments of
the corresponding measure:
\begin{equation}\label{1}
P_q=\sum_{i=1}^M\, p_i^q=\int\, M^{-q\alpha} \rho(\alpha)\,
d\alpha\,.
\end{equation}
Substituting in the above definition the relation
$\rho(\alpha)\sim M^{f(\alpha)}$ one can evaluate the integral in
the large-$M$ limit by saddle-point method. One then finds the
relation between $\tau_q$ and $f(\alpha)$ given by the Legendre
transform:
\begin{equation}\label{1a}
P_q\sim M^{-\tau_q},\quad \tau_q=q\alpha-f(\alpha) \quad
\mbox{where}\quad q = \frac{df}{d\alpha}\,\,.
\end{equation}
The relation is valid as long as the $\alpha-$integral is
dominated by the saddle point. It is easy to see however that at
large enough $|q|$ the integral should be dominated rather by the
vicinity of the thresholds $\alpha_{\pm}$ resulting in linear
behaviour of the exponents with $q$, i.e. $\tau_q=q\alpha_\pm$.

The above description is valid for multifractal measures of any
nature. In recent years important insights were obtained for
disorder-generated multifractality, see \cite{ME} and also
\cite{FRL} for a comprehensive discussion in the context of
Anderson localisation transitions, and \cite{MG} for an example
related to polymers in disordered media. One of specific features
of multifractality in the presence of disorder is a possibility of
existence of two different sets of exponents, $\tau_q$ versus
$\tilde{\tau}_q$, governing the scaling behaviour of typical $P_q$
versus disorder averaged IPR's, $<P_q>\sim M^{-\tilde{\tau}_q}$.
Here and henceforth the brackets stand for the averaging over
different realisations of the disorder. The difference is related
to a possibility of disorder-averaged moments to be dominated by
exponentially rare configurations. A related aspect of the problem
is that the "annealed" multifractality spectrum
 recovered from the disorder-averaged
multifractal exponents $\tilde{\tau}_q$ via the Legendre transform
(\ref{1}) can be negative: $\tilde{f}(\alpha)<0$. Indeed, those
values reflect events which are exponentially rare \cite{negf} and
need exponentially many realisations of disorder to be observed
experimentally or numerically. In the context of Anderson
localisation the disorder-averaged moments of wavefunction
intensities are readily available via standard techniques in the
non-linear $\sigma-$model framework, see \cite{ME,FRL} and
references therein. At the same time extracting typical values of
the multifractality exponents in that case is a much more
challenging task which has been successfully accomplished only
very recently \cite{FRL}. In the present paper we would like to
concentrate on a different type of models where, in contrast,
calculating disorder-averaged moments in the full parameter range
is more difficult, whereas the typical values of IPR exponents are
readily accessible.

Arguably the simplest model with disorder-induced multifractality
which attracted considerable interest in recent years is the case
 of a single classical particle subject to a random Gaussian
potential $V({\bf x})$ logarithmically correlated in space:
\begin{equation}\label{2}
\left\langle V\left({\bf x}_1\right) \, V\left({\bf
x}_2\right)\right\rangle=-\,g^2\ln{\left[\frac{({\bf x}_1-{\bf
x}_2)^2+a^2}{L^2}\right]},\quad a\ll L, \quad {\bf x}\in
\mathbb{R}^N\,,
\end{equation}
where we assume $|{\bf x}|<L$, and the parameter $a$ stands for a
small-scale cutoff. From the point of view of equilibrium
statistical mechanics the model is characterized by the
Gibbs-Boltzmann measure $p_{\beta}({\bf
x})=\frac{1}{Z(\beta)}\exp{-\beta V({\bf x})}\,$ as a function of
the inverse temperature $\beta={1}/{T}$, and the sample size $L$.
The normalization $\int_{|{\bf x}|\le L} p_{\beta}({\bf x}) d {\bf
x}\,=1$ implies the value of the partition function to be given by
\begin{equation}\label{freeendef}
Z(\beta)=\int_{|{\bf x}|\le L} \exp{-\beta V({\bf x})}\, d {\bf
x}\,.
\end{equation}
According to the general discussion, the multifractal structure of
the Gibbs-Boltzmann measure can be extracted from the knowledge of
moments
\begin{equation}\label{BGmom}
\quad P_q=\int_{|{\bf x}|\le L} p^q_{\beta}({\bf x})\, d {\bf
x}=\frac{Z(\beta q)}{\left[Z(\beta)\right]^q}\sim
L^{-N\tau_q}\quad \mbox{as}\quad L\to \infty\,.
\end{equation}
Identifying $M\sim (L/a)^N$ , the Eqs.(\ref{BGmom}) and
(\ref{freeendef}) imply the following expression for the {\it
typical} exponents $\tau_q$ in terms of the appropriately
normalized free energy of the system
\begin{equation}\label{BGmom2}
\quad \tau_q=|q|\beta{\cal F}(|q|\beta)-q\beta{\cal F}(\beta),
\quad {\cal F}(\beta)=-\lim_{M\to
\infty}\frac{\left\langle\ln{Z(\beta)}\right\rangle}{\beta
\ln{M}}\,.
\end{equation}

Although the model is well-defined in any $N-$dimensional space
\cite{CLD,FB1} it is two-dimensional situation which attracts most
attention, with Eq.(\ref{2}) describing the correlations of the
Gaussian free field. In particular, for $N=2$ the corresponding
statistics of Gibbs-Boltzmann weights is known to be deeply
related to a variety of interesting physical problems, ranging
from quantum mechanics of Dirac particles in a random magnetic
field \cite{LFSG,2d} to Liouville model of quantum
gravity\cite{Liouville} and theory of self-gravitating particles
\cite{selfgrav}, see a detailed discussion in \cite{CLD}.
Actually, the dimensionality of space plays in many respects only
secondary role and many (although not all) essential features of
the model are expected to be $N-$independent. The latter point of
view is amply supported by the renormalization group arguments
\cite{CLD} and by explicit computations in $N=\infty$ \cite{FB1}
and $N=1$ \cite{FB2}. It is also worth mentioning that various
one-dimensional versions of the problem attracted considerable
interest recently in the context of multifractal random
walks\cite{BMD,Ostrov}, extreme value
statistics\cite{FB2,CLD,FLDR} and quantum gravity-related
\cite{DS} probabilistic questions, see \cite{Vargas} and
references therein.

\section{ Random Energy Model as a toy model for disorder-induced multifractality}

A particular extreme "toy model" case of the problem is
represented by the famous Random Energy Model (REM) by Derrida
\cite{REM,GD89}, which amounts to replacing random potential by a
collection of $M$ independent Gaussian variables, after the
natural identification $M\sim (L/a)^N$ and with the variances
scaled with $M$ in the same way as in the logarithmic case:
$<V_i^2>=2g^2\ln{M}$. The only control parameter for the model is
$\gamma=\beta^2g^2$, and REM is simple enough to allow explicit
calculation of the free energy \cite{REM,GD89}. The typical
multifractality exponents turned out to be given by \cite{2d}
\begin{equation}\label{typ}
\tau_{q>1}(\gamma)=\left\{\begin{array}{c}(q-1)(1-\gamma q), \quad 0\le \gamma<\frac{1}{q^2}\\
q(1-\sqrt{\gamma})^2, \quad \frac{1}{q^2}<\gamma<1\,\\
0,\,\,\, \quad \gamma>1\,
\end{array}\right.\,.
\end{equation}
 The phenomenon of vanishing of the exponents  $\tau_{q>1}$ in the low-temperature phase $\gamma>1$
  is called {\it freezing} and is qualitatively interpreted in terms of
the Boltzmann measure being essentially localised on a few sites
for low enough temperature or strong enough disorder.  The typical
multifractality spectrum corresponding to the above exponents is
\cite{2d}
\begin{equation}\label{typspectrum}
f(\alpha)=\left\{\begin{array}{c}1-\frac{1}{4\gamma}\left[\alpha-(1+\gamma)\right]^2\quad
 \mbox{for}\quad  \gamma<1\\ -\frac{1}{4\gamma}\left[\alpha^2-4\sqrt{\gamma}\alpha\right]\quad
 \mbox{for} \quad \gamma>1
\end{array}\right.\,,
\end{equation}
where the expression in the first line assumes the range of
exponents $\alpha_{-}=(1-\sqrt{\gamma})^2\le \alpha\le
 1+\gamma$, whereas in second line $0\le \alpha\le
 2\sqrt{\gamma}$. Thinking in terms of the multifractality spectrum
 it is  easy to see that the freezing
phenomenon at $\gamma>1$ is related to $\alpha_{-}=0$, when the
leftmost end of the curve $f(\alpha)$ hits the vertical axis
precisely at zero level: $f(0)=0$. Similarly, the change of
behaviour of the typical exponent $\tau_q$ for $\gamma>1/q^2$ is
induced by dominance of the point $\alpha_{-}$ in the integration
over $\alpha$, Eq.(\ref{1}).

Although obtained in the framework of REM approximation, the above
features of the typical spectrum are expected to be shared by all
the logarithmic model for any $N\ge 1$ \cite{2d,CLD}, which is
indeed confirmed by explicit calculations on $N=\infty$ \cite{FB1}
and $N=1$ case \cite{FB2}. In contrast to the case of typical
exponents $\tau_q$, extracting the "annealed" exponents
$\tilde{\tau}_q$ from disorder-averaged moments in the logarithmic
models poses a serious technical challenge. The only systematic
attempt in this direction was undertaken for $N=2$ in the
framework of mapping to the Liouville model of quantum
gravity\cite{Liouville} where it was found that
$\tilde{\tau}_{q>1}=(q-1)(1-\gamma q)$ for $0\le
\gamma<\gamma_q=1/(2q-1)<1$. However, for $\gamma>\gamma_q$ the
Liouville theory was observed to develop unsurmountable
singularities and yielded no reliable value of the exponents
$\tilde{\tau_q}$. This state of matter clearly calls for
reconsidering the problem within the general framework of
logarithmic models.

The natural starting point is again the standard REM representing
in many respects a zero-dimensional limit of the logarithmic
models. For such a "toy model" case the disorder-averaged moments
(IPR's) $<P_q>$ can be evaluated by a well-controlled
methods\cite{GD89,DT}. Surprisingly, explicit expressions for the
IPR's seem to be available in the literature
 only in the low-temperature phase $\gamma>1$ \cite{DT,Derrida3,BM}. Extending
the analysis of \cite{GD89,DT} we find for $M\gg 1$ and $q>1$:

\begin{equation}\label{REM}
\langle P_q \rangle=\left\langle Z(\beta
q)/\left[Z(\beta)\right]^q\right \rangle =
\left\{\begin{array}{c}M^{-(q-1)(1-\gamma q)}, \quad 0\le
\gamma<1/(2q-1)\\
\frac{M^{-\frac{(1-\gamma)^2}{4\gamma}}}{2\sqrt{\pi\gamma\ln{M}}}
\frac{\Gamma\left(\frac{1+\gamma}{2\gamma}\right)\Gamma
\left(q-\frac{1+\gamma}{2\gamma}\right)}{\Gamma(q)}, \quad \frac{1}{2q-1}<\gamma<1\\
\\ \frac{\Gamma\left(q-\frac{1}{\sqrt{\gamma}}\right)}{\Gamma(q)
\Gamma\left(1-\frac{1}{\sqrt{\gamma}}\right)}, \quad
\gamma>1\end{array}\right.
\end{equation}
where we included in the last line the well-known low-temperature
results of \cite{DT,Derrida3,BM}, with $\Gamma(x)$ standing for
the Euler gamma-function. The fact that for $\gamma>1$ the moments
remain finite in the limit $M\to \infty$ reflects the
quasi-localised nature of the Boltzmann-Gibbs measure in the
low-temperature phase. As a consequence, the "annealed"
multifractal exponents remain frozen: $\tilde{\tau_q}=0$. In the
high-temperature phase the exponents $\tilde{\tau_q}$ are
non-vanishing and in the range $0<\gamma<\frac{1}{q^2}$ typical
and annealed exponents coincide. The annealed exponents actually
keep that common value up to $\gamma=\gamma_q=1/(2q-1)$. Both the
value $\tilde{\tau}_q$ for $\gamma<\gamma_q$ and the value of the
threshold $\gamma_q$ are in full agreement with the Liouville
model analysis \cite{Liouville}. Finally, in the range
$\gamma_q<\gamma<1$ the annealed exponents change drastically by
acquiring the $q-$independent value which relates to $\gamma$ in a
non-polynomial way $\tilde{\tau}_q=\frac{(1-\gamma)^2}{2\gamma}$.
The behaviour of typical and annealed multifractal exponents for
various values of the control parameter $\gamma$ is summarized in
the diagram, and is further discussed below.
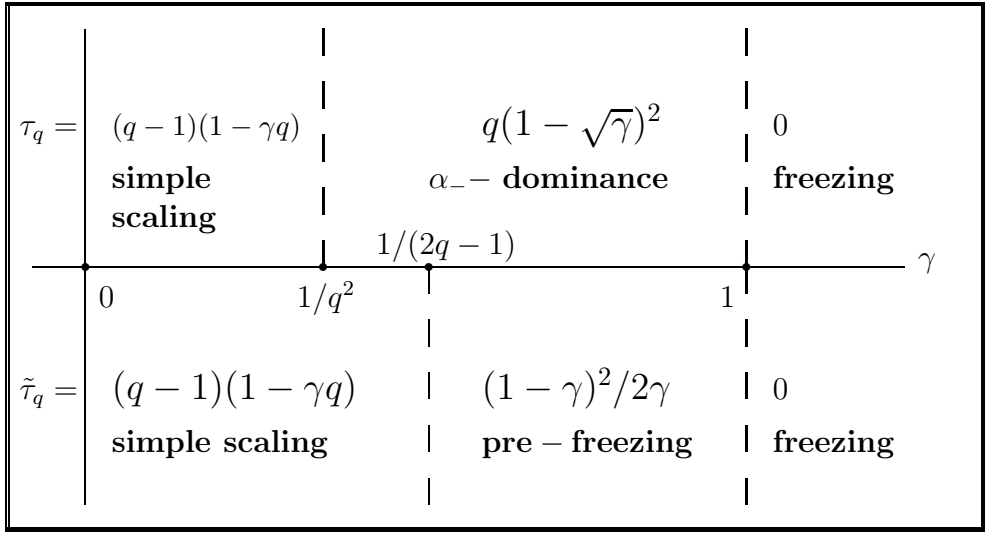
\begin{figure}[h!]
%\begin{center}
\begin{picture}(200,210)
\put(5,50){$ \tilde{\tau}_q=$} \put(5,150){$\tau_q=$}
\put(40,130){\bf simple}\put(40,115){\bf scaling}
\put(40,50){\large$(q-1)(1-\gamma q)$}
\put(180,50){\large$(1-\gamma)^2/2\gamma$}\put(180,150){\large$q(1-\sqrt{\gamma})^2$}
\put(160,130){$\alpha_{-}-$ {\bf dominance}} \put(290,50){$0$}
\put(40,30){\bf simple scaling}\put(180,30){${\bf
pre-freezing}$}\put(290,30){${\bf freezing}$}

\put(290,150){$0$}\put(290,130){${\bf freezing}$}
 \put(30,10){\line(0,10){180}}\put(30,100){\circle*{3}}\put(35,85){$0$}
\put(40,150){\small$(q-1)(1-\gamma q)$}
\put(10,100){\line(1,0){330}}\put(345,100){${\mathbf{\gamma}}$}
\put(120,100){\circle*{3}}\put(160,100){\circle*{3}}\put(280,100){\circle*{3}}
\put(270,85){$1$}\put(110,85){$1/q^2$}\put(140,105){$1/(2q-1)$}
\multiput(280,100)(0,20){5}{\line(0,1){10}}
\multiput(280,10)(0,20){5}{\line(0,1){10}}
\multiput(120,100)(0,20){5}{\line(0,1){10}}
\multiput(160,10)(0,20){5}{\line(0,1){10}}

\put(0,200){\line(1,0){370}}\put(0,199){\line(1,0){370}}\put(0,0){\line(1,0){370}}
\put(0,1){\line(1,0){370}}\put(0,0){\line(0,1){200}}\put(1,0){\line(0,1){200}}
\put(370,0){\line(0,1){200}}\put(369,0){\line(0,1){200}}

\end{picture}
\caption{Regimes of behaviour for two types of multifractal
exponents for the Random Energy Model: typical exponent $\tau_q$ vs. annealed $\tilde{\tau}_q$  at different values of effective disorder $\gamma$.}
%\end{center}
\end{figure}

Using Eq.(\ref{REM}) one can restore the corresponding mean
density of multifractal exponents:
$$<\rho(\alpha)>=\left\langle\sum_{i=1}^M\,\delta(\alpha-\alpha_i)\right\rangle
\approx C(M,\alpha)\,M^{\tilde{f}(\alpha)}$$ where the "annealed"
multifractality spectrum turned out to be given by
\begin{equation}\label{spectrum}
\tilde{f}(\alpha)=\left\{\begin{array}{c}1-\frac{1}{4\gamma}\left[\alpha-(1+\gamma)\right]^2,\quad
 \mbox{for} \quad 0\le \gamma<1 \quad \mbox{and}\quad  0\le \alpha\le
 1+\gamma,\\ -\frac{1}{4\gamma}\left[\alpha^2-4\sqrt{\gamma}\alpha\right]\quad
 \mbox{for}\quad  \gamma>1 \quad \mbox{and} \quad 0\le \alpha\le
 2\sqrt{\gamma}
\end{array}\right.\,.
\end{equation}
These expressions show that the disorder-averaged (or "annealed")
spectrum is precisely the same as the typical spectrum
corresponding to Eq.(\ref{typspectrum}), with the only essential
difference being that annealed spectrum in the first line of
Eq.(\ref{spectrum}) becomes negative in the range $0\le
\alpha<\alpha_{-}=(1-\sqrt{\gamma})^2$. These values of $\alpha$
correspond to exponentially rare events in full agreement with our
earlier discussion and the picture developed in \cite{ME,negf}.
The value $\alpha=0$ is the lowest possible value of the exponent
$\alpha$, and for this reason is frequently called the
"termination point" of the (disorder-averaged) multifractality
spectrum\cite{ME}. Further substituting the corresponding
$<\rho(\alpha)>$ to the integral over $\alpha$\footnote{It is
worth mentioning that for REM one can extract not only the
multifractality spectrum, but also the expressions for the
pre-exponential factors $C(M,\alpha)$ which goes beyond the
standard precision of the multifractality analysis. For example,
in the high-temperature phase $ 0\le \gamma<1$ one finds $
C(M,\alpha)=\frac{1}{2}\sqrt{\frac{\ln{M}}{\pi\gamma}}
\left(1-M^{-\alpha}\right)^{-\frac{\alpha-1-\gamma}{2\gamma}}$.},
see Eq.(\ref{1}), we find that in the range
$\frac{1}{2q-1}<\gamma<1$ the integral for $\left\langle
P_{q>1}\right\rangle$ is actually dominated in the limit $M\gg 1$
by the vicinity of the lower integration limit $\alpha=0$ rather
than by the stationary point of the integrand. It is in this way
that $\tilde{\tau}_q$ acquires the q-independent value
$\frac{(1-\gamma)^2}{2\gamma}$, cf. Eq.(\ref{REM}), as was indeed
anticipated in \cite{ME} on heuristic grounds. We see that the
actual mechanism behind the observed drastic change of the
annealed exponents is very close to one forcing the exponent
freezing at $\gamma>1$, and in a sense is the precursor of the
latter behaviour. For this reason it is natural to suggest to call
this phenomenon pre-freezing. We shall also see later on that the
replica approach gives additional support to associating the
observed behaviour with a partial freezing of certain kind.

\section{Annealed multifractality exponents for the circular logarithmic model}

After the detailed understanding of the toy REM limit it is
natural to try to extract annealed exponents for a more realistic
case of one-dimensional model with logarithmic correlations. The
most promising instance of 1D system of that type is arguably the
"circular logarithmic model" (CLM) introduced in \cite{FB2}. To
define it, consider the lattice of $M$ points positioned
equidistantly at the circumference of a circle of a radius $a\ll
R<L$ in the standard two-dimensional free field with correlations
Eq.(\ref{2}). The angular coordinates of the points are given by
$\theta_k=\frac{2\pi}{M}k$, $k=1,2,\ldots,M$. Then it is easy to
to see that the values of the free field $V_i$ associated with
those points are characterized by covariances \be \label{circular}
C_{kl}=\langle
V_kV_l\rangle=-g^2\ln{\left\{4\sin^2\frac{\theta_k-\theta_l}{2}\right\}}+2g^2\ln{(L/R)},
\quad k\ne l  \label{1b}\ee The first term in (\ref{circular})
defines precisely CLM as described in \cite{FB2}, with constant
second term playing a somewhat trivial role and thus omitted
henceforth. It turns out that the consistency of the procedure
requires to choose the variance $\langle V_k^2\rangle=V^2$ to
satisfy $V^2\ge 2g^2\ln{M}$, and following \cite{FB2} we choose
simply $V^2= 2g^2\ln{M}$. The partition function for our model is
defined in the standard way through
$Z_{\beta}=\sum_{i=1}^Me^{-\beta V_i}$, with the goal to evaluate
the IPR moments $P_q=\left\langle Z(\beta
q)/\left[Z(\beta)\right]^q\right \rangle$ in the limit $M\gg 1$
for various values of the control parameter $\gamma=\beta^2g^2$.

 In the rest of the paper we restrict our explicit
calculations by the simplest representative case $q=2$, and employ
a variant of the replica trick:
\begin{equation}\label{replica}
\langle P_2\rangle=\left\langle
Z(2\beta)/\left[Z(\beta)\right]^2\right \rangle= \lim_{n\to 0
}\left\langle Z(2\beta) \left[Z(\beta)\right]^{n-2}\right \rangle,
\end{equation}
implying a kind of continuation from integer values $n\ge 2$ to
$n=0$. The key point is that for integer $n\ge 2$ the disorder
averaging in (\ref{replica}) can be performed by the method
described in detail in \cite{FB2}, which gives, in particular
\begin{equation}\label{replica1a}
\left\langle Z(2\beta) \left[Z(\beta)\right]^k\right
\rangle|_{M\gg 1} =M^{1+k+\gamma(k+4)}J_k(\gamma), \quad 0\le
k<\frac{1}{\gamma}-3
\end{equation}
where $J_k(\gamma)$ is the Dyson-Morris Integral \cite{Forrester}
 given by a product of gamma-functions:
\begin{eqnarray}\label{DM} {\cal
J}_k(\gamma)&=&\frac{1}{(2\pi)^k} \int_0^{2\pi}\,d\theta_1 \ldots
\int_{0}^{2\pi}\,d\theta_k \prod_{p<q}^k
\left|e^{i\theta_p}-e^{i\theta_q}\right|^{-2\gamma}\prod_{l=1}^k\,
\left|1-e^{i\theta_l}\right|^{-4\gamma}\\
&=&
\frac{1}{[\Gamma(1-\gamma)]^{k-1}}\frac{\Gamma[1-\gamma(k+2)]\Gamma[1-\gamma(k+3)]}
{\Gamma[1-\gamma(k+1)]\Gamma(1-2\gamma)\Gamma(1-3\gamma)}\,\,.
\end{eqnarray}
Performing in the above expression the "naive" replica limit $k\to
-2$ we arrive at the following expression for the IPR:
\begin{equation}\label{replica1}
\langle
P_2\rangle=M^{-(1-2\gamma)}\frac{[\Gamma(1-\gamma)]^{4}}{\Gamma(1+\gamma)\Gamma(1-2\gamma)\Gamma(1-3\gamma)}\,\,.
\end{equation}
We see that the value of the annealed multifractality exponent
$\tilde{\tau}_2=1-2\gamma$ which emerges from our calculation
coincides with the "simple scaling" value
$\tilde{\tau}_q=(q-1)(1-q\gamma),\,\, q=2$ discussed by us
earlier. Moreover, the consistency of the above procedure
obviously requires $0\le \gamma<1/3$, with the upper limit being
precisely the threshold $\gamma_{q=2}=1/3$ of validity of the
above "simple scaling" regime. We conclude that a simple-minded
replica limit could be employed to produce meaningful results only
 as long as the pre-freezing phenomenon responsible for the change of
multifractality exponent is not operative.

\section{Infinite-dimensional limit: prefreezing via replica symmetry breaking}

To get some progress in understanding of the mechanisms behind the
failure of the simple scaling in replica approach, we turn from
now on to another exactly solvable limit of the logarithmic model,
that is to the infinite-dimensional case. The free energy, hence
the typical multifractality spectrum, was calculated in \cite{FB1}
in the framework of the replica trick, and very recently confirmed
by rigorous mathematical methods \cite{AK}. The system was found to
display the REM-type freezing transition at $\gamma=\beta^2g^2=1$,
with the low-temperature phase $\gamma>1$ described by the
standard one-step replica symmetry breaking pattern. The meaning
of the freezing could be elucidated by invoking the probability
for two independent particles distributed in such random potential
according to the Boltzmann-Gibbs measure to end up at a distance
of order of the small cutoff scale $a^2$. The probability was
found \cite{FB1} to tend to zero in the thermodynamic limit
$L/a\to \infty$ everywhere in the high-temperature phase $0\le
\gamma<1$, confirming the particle delocalization over the sample.
In contrast, in the lower temperature phase $\gamma>1$ two
particles have a finite probability to be trapped at the
small-scale distance even in the infinite sample.

Our starting point here is again the identity (\ref{replica}).
Employing it one can easily perform the disorder average for any
integer number of replica $n\ge 2$. After appropriate rescaling of
the coupling constant $g\to g\sqrt{N}$ and length scales $L\to
L\sqrt{2N}$ and $a\to a\sqrt{2N}$ the manipulations similar to
those described in detail in \cite{FB1} yield a convenient
representation for the IPR in terms of an integral over a positive
definite matrix $\mathbf{Q}$ of the size $(n-1)\times (n-1)$ with
entries $q_{\mu,\nu}$. We have
\begin{equation}\label{inf1}
\langle P_2\rangle=\lim_{n\to 0}{\cal C}_{N,n}(a)L^{\gamma n^2}\,
\int_{D_\mathbf{Q}} \left(\mbox{det}\mathbf{Q}\right)^{-n/2} e^{-
N\Phi_n (\mathbf{Q}) }\, d\mathbf{Q}
\end{equation}
where
\[
 \Phi_n (\mathbf{Q})=-
 \frac{1}{2}\ln{(\det{\mathbf{Q}})}+\gamma\sum_{1\le \mu<\nu\le n-2}
\ln\left[\frac{1}{2}(q_{\mu,\mu}+q_{\nu,\nu})-q_{\mu,\nu}+a^2\right]+
\]
\begin{equation}\label{inf2}
+2\gamma\,\sum_{\mu=1}^{n-2}\ln
\left[\frac{1}{2}(q_{\mu,\mu}+q_{n-1,n-1})-q_{\mu,n-1}+a^2\right],\quad
\gamma=\beta^2g^2,
\end{equation}
and  the integration domain $D_Q$ in the above expression is
simply $D_Q=\{\mathbf{Q}>0,\, q_{\mu,\mu}\le \,L^2,\, \mu=1,\ldots
n-1\}$. The proportionality constant ${\cal C}_{N,n}(a)$ is also
explicitly known, but its value is inessential for the subsequent
calculation.

The shape of the integrand in (\ref{inf1}) is suggestive of
application of the saddle-point method for evaluation of the
integral in the large-$N$ limit. The corresponding saddle-point
equations for the entries of the matrix $\mathbf{Q}$ amount to
$\partial \Phi/\partial{q_{\mu,\nu}}=0$ for any choice of the
indices $1\le \mu\le \nu\le (n-1)$. A closer inspection of the
replica limit $n\to 0$ reveals however that solutions to the
saddle-point equations do not actually exist unless one fixes all
the diagonal entries $q_{\mu,\mu}, \,\, 1\le \mu\le (n-1)$ of the
matrix $\mathbf{Q}$ to the boundary of the integration domain by
setting $q_{\mu,\mu}=L^2$, and excluding them from the variational
procedure ( cf. a similar result in \cite{FB1}). The remaining
off-diagonal entries should be found from the system of equations:
\begin{equation}\label{inf0a}
\left[\mathbf{Q}^{-1}\right]_{\mu,\nu}+\gamma\,\frac{1}{L^2-q_{\mu,\nu}+a^2}=0,
\quad 1\le \mu<\nu\le n-2
\end{equation}
\begin{equation}\label{inf0b}
\left[\mathbf{Q}^{-1}\right]_{\mu,n-1}+2\gamma\,\frac{1}{L^2-q_{\mu,n-1}+a^2}=0,
\quad  1\le \mu\le n-2
\end{equation}

The apparently non-equivalent roles played by the special replica
index $\nu=n-1$ and the rest of $n-2$ replicas  in the above
equations are inherited from the structure of IPR representation
in the replica method, Eq.(\ref{replica}). The only ansatz for $Q$
respecting the full permutation symmetry between the $(n-2)$
equivalent replicas is given by
\begin{equation}\label{inf3}
\mathbf{Q}_{r.s.}=\left(\begin{array}{cc} \mathbf{Q}^{(0)} & \mathbf{v}^T\\
\mathbf{v} & L^2 \end{array} \right), \quad
\mathbf{Q^{(0)}}_{\mu,\nu}=(L^2-q_0)\delta_{\mu,\nu}+q_{0}, \quad
1\le \mu,\nu\le n-2\,
\end{equation}
where $\mathbf{Q}^{(0)}>0$ is of the size $(n-2)\times (n-2)$, and
the $(n-2)-$component vector $\mathbf{v}$ is of the form
$\mathbf{v}=v(1,\ldots,1)$. Two parameters $q_0$ and $v$
characterising such a replica-symmetric solution satisfy in the
limit $n\to 0$ the system of equations
\begin{equation}\label{inf4}
\frac{q_0-v^2/L^2}{(L^2-q_0)(L^2-3q_0+2v^2/L^2)}=\frac{\gamma}{(L^2-q_0+a^2)}\,,
\end{equation}
\begin{equation}\label{inf4a}
\frac{v}{L^2(L^2-3q_0+2v^2/L^2)}=\frac{2\gamma}{(L^2-v+a^2)}\,.
\end{equation}
The only solution of such a system existing in the
high-temperature phase in the thermodynamic limit $L\gg a$ has the
following form
\begin{equation}\label{inf5}
 q_0 = \frac{\gamma(1+3\gamma)}{(1+\gamma)^2}\,L^2+O(a^2), \quad
v= \frac{2\gamma}{1+\gamma}\,L^2+O(a^2) \,.
\end{equation}
The condition $q_0<L^2$ which is readily seen to be satisfied
everywhere in the high-temperature phase $0\le \gamma < 1$ ensures
that $\mathbf{Q^{(0)}}>0$.

Substituting these values back to the relation (\ref{inf2}) and
using the identity
\begin{equation}
\det{\mathbf{Q}_{r.s.}}=L^2(L^2-q_0)^{n-3}\left[L^2-v^2/L^2+(n-3)(q_0-v^2/L^2)\right]
\end{equation}
we find to the leading order $ \Phi_{n\to 0}
(\mathbf{Q}_{r.s.})=(1-2\gamma)\ln{L}$ which in turn yields the
asymptotic behaviour for the averaged inverse participation ratio
$\langle P_2\rangle\propto e^{-N\Phi_{n\to 0}}\propto
M^{-(1-2\gamma)}$ where we have identified $M\sim L^N$. We arrive
to an important conclusion of the replica symmetric ansatz
Eq.(\ref{inf3}) implying necessarily the "simple scaling" result
$\tilde{\tau}_2=1-2\gamma$ for the annealed multifractality
exponent, see the discussion after (\ref{replica}). As is clear,
the latter value  cannot have any meaning at least for
$\gamma>1/2$, which in turn implies that the replica-symmetric
solution of the problem cannot be valid in the whole
high-temperature phase $0\le \gamma<1$. The only way out is
therefore to look for an appropriate scheme of replica symmetry
breaking which occurs already in the high-temperature phase.

To get a guiding idea in our search for a solution to the
saddle-point equations (\ref{inf0a},\ref{inf0b}) which goes beyond
the replica symmetric ansatz, eq.(\ref{inf3}), it is useful to
recall that the singled-out replica indexed with $\nu=n-1$ has its
origin in representing the factor $Z(2\beta)$ in the averaged
inverse participation ratio, cf. Eq.(\ref{replica}). As such that
special replica is effectively "colder" then all other equivalent
replica indexed with $1,2,\ldots, n-2$ which originated from the
factors $Z(\beta)$. It is then natural to suppose that by
interacting with other replica the colder one could eventually
"pre-freeze" a certain group of replica around it. The
manifestations of the freezing mechanism within the replica
approach was discussed for the present model in detail in
\cite{FB1}. Employing it one should expect such a "pre-freezing"
to manifest itself via existence of a diagonal block inside the
matrix $\hat{\mathbf{Q}}$ having in the thermodynamic limit $L\gg
a$ all entries equal , up to the leading order, to $L^2$.

The simplest ansatz for the matrix $\mathbf{Q}$ compatible with a
possibility of such pre-frozen group of $m$ replicas, $0\le m\le
n-2$, would be of the following structure:
\begin{equation}\label{inf6}
\fl \mathbf{Q}_{r.s.b}=\left(\begin{array}{ccc} \mathbf{Q}^{(1)} &q_3 \mathbf{E}^T& \mathbf{v}_1^T\\
q_3\mathbf{E}&\mathbf{Q}^{(2)}&\mathbf{v}^T_2\\
\mathbf{v}_1 &\mathbf{v}_2 & L^2 \end{array} \right), \quad
\begin{array}{l}
\mathbf{Q^{(1)}}_{\mu,\nu}=(L^2-q_1)\delta_{\mu,\nu}+q_{1}, \quad
1\le \mu,\nu\le n-m-2\,\\
\mathbf{Q^{(2)}}_{\mu,\nu}=(L^2-q_2)\delta_{\mu,\nu}+q_{2}, \quad
1\le \mu,\nu\le m\end{array}
\end{equation}
where all entries of the matrix $\mathbf{E}$ of the size $m\times
(n-m-2)$ are equal to $1$, the $(n-m-2)-$component vector
$\mathbf{v}_1$ has all components equal to $v_1$ and $m-$component
vector $\mathbf{v}_2$ is of the same structure:
$\mathbf{v}_2=v_2(1,\ldots,1)$. The values of five parameters
$q_1,q_2,q_3,v_1,v_2$ for a given size $m$ of the pre-frozen block
are to be found from the saddle-point conditions (\ref{inf0a}),
(\ref{inf0b}) in the replica limit $n\to 0$. Finally, the
parameter $m$ satisfying in the replica limit inequality $-2\le
m\le 0$ should itself be chosen as to extremize the resulting
$\Phi(\mathbf{Q})$, eq.(\ref{inf2}), which is evaluated with the
help of the identity
\begin{eqnarray}\label{inf9}
\fl \nonumber &&
\det{\hat{\mathbf{Q}}_{r.s.b.}}=L^2(L^2-q_1)^{n-m-3}\left[L^2-v_1^2/L^2+(n-m-3)(q_1-v_1^2/L^2)\right]\\
\fl && \times
(L^2-q_2)^{m-1}\left[L^2-v_2^2/L^2+(m-1)(q_2-v_2^2/L^2)-r\,m(q_3-v_1v_2/L^2)\right]\\
&& \nonumber \mbox{where} \quad
r=\frac{n-m-2}{L^2-v_1^2/L^2+(n-m-3)(q_1-v_1^2/L^2)}\,.
\end{eqnarray}

Substituting the Ansatz (\ref{inf6}) into the saddle-point
equations (\ref{inf0a}), (\ref{inf0b}) one arrives in the limit
$n=0$ after straightforward but lengthy algebraic manipulations to
the set of equations:
\begin{equation}\label{inf7a}
 \frac{q_1}{L^2-q_1}=\gamma\,\frac{L^2-q_1(m+3)}{L^2-q_1+a^2}+
\gamma\,m\,\frac{q_3}{L^2-q_3+a^2}+2\gamma\,\frac{v_1}{L^2-v_1+a^2}
\end{equation}
\begin{equation}\label{inf7b}
 \frac{q_3}{L^2-q_2}=\gamma\,\frac{L^2-q_1(m+3)}{L^2-q_3+a^2}+
\gamma\,m\,\frac{q_3}{L^2-q_2+a^2}+2\gamma\,\frac{v_1}{L^2-v_2+a^2}
\end{equation}
\begin{equation}\label{inf7c}
 \frac{q_3}{L^2-q_1}=\gamma\,\frac{L^2+q_2(m-1)}{L^2-q_3+a^2}-
\gamma\,(m+2)\,\frac{q_3}{L^2-q_1+a^2}+2\gamma\,\frac{v_2}{L^2-v_1+a^2}
\end{equation}
\begin{equation}\label{inf7d}
 \frac{q_2}{L^2-q_2}=\gamma\,\frac{L^2+q_2(m-1)}{L^2-q_2+a^2}-
\gamma\,(m+2)\,\frac{q_3}{L^2-q_3+a^2}+2\gamma\,\frac{v_2}{L^2-v_2+a^2}
\end{equation}
\begin{equation}\label{inf7e}
 D\,v_1=2\gamma\,\frac{L^2-q_1(m+3)}{L^2-v_1+a^2}+2\gamma\,m\frac{q_3}{L^2-v_2+a^2}
\end{equation}
\begin{equation}\label{inf7f}
 1=D\,L^2+2\gamma\,(m+2)\frac{v_1}{L^2-v_1+a^2}-2\gamma\,m\frac{v_2}{L^2-v_2+a^2}\,,
\end{equation}
where we found it convenient to use the parameter $D\equiv
[\mathbf{Q}^{-1}]_{n-1,n-1}$.

It is immediate to check that the above system of equations always
has the replica symmetric type of solution such that
$q_1=q_2=q_3\equiv q_0$ and $v_1=v_2\equiv v$. Indeed, such a
substitution results in $m$ dropping out from the equations, and
$q_0$ and $v$ satisfying equations (\ref{inf4}).  At the same time
there exists another type of solution which explicitly depends on
$m$ and is given in the limit $L\gg a$ by
\begin{eqnarray}\label{inf8}
\nonumber \fl &&
q_1=\frac{\gamma}{(1+\gamma)^2}\left[1+\gamma(m^2+3m+3)\right]\,L^2+O(a^2),
 \quad q_2=L^2+O(a^2)\\
\fl && q_3=\frac{\gamma}{1+\gamma}\,(m+2)\,L^2+O(a^2), \,\,
v_1=\frac{\gamma}{1+\gamma}\,(m+2)\,L^2+O(a^2), \quad v_2
=L^2+O(a^2).
\end{eqnarray}
We indeed see that all the off-diagonal entries inside the block
$\left(\begin{array}{cc}\mathbf{Q}^{(2)}&\mathbf{v}^T_2\\
\mathbf{v}_2 & L^2 \end{array}\right)$ are such that the (squared)
"distances" $d=q_2-L^2$ and $\tilde{d}=v-L^2$ are of the order of
the cut-off scale $a^2$. The latter feature is precisely the
property of being frozen, see \cite{FB1} for a more detailed
discussion. It is also worth noting that though in the limit $m\to
0$ in (\ref{inf8}) $q_1$ and $v_1$ tend to the replica symmetric
values, they remain different from $q_2,q_3$ and $v_2$,
respectively, so that even in the limit $m\to 0$ the new solution
does not tend to the replica-symmetric one.

Substituting now the solution (\ref{inf8}) into (\ref{inf9}) one
finds after straightforward algebraic manipulations
$\frac{1}{2}\ln{\det{\mathbf{Q}_{r.s.b.}}}=-(1+m)\ln{L}+O(\ln{a})$,
which after more algebra results in the value of the functional
$\Phi_{n\to 0}(\mathbf{Q})$ for a given value of the parameter
$m$:
\begin{equation}\label{inf10}
\Phi_{n\to
0}(\mathbf{Q}_{r.s.b})=\ln{L}\left(1+m-\gamma[m^2+3m+2]\right)+O(\ln{a})
\end{equation}
Extremizing this functional with respect to $m\in [-2,0]$ one
immediately finds that for $1/3<\gamma<1$ the extremum is at an
internal point of the interval
$m=m_*=\frac{1}{2}\frac{1-3\gamma}{\gamma}$, whereas for $0\le
\gamma\le 1/3$ the extremum is at the boundary point $m=0$.
Substituting those values to the functional leads in the
thermodynamic limit to value $\Phi_{n\to 0}(\mathbf{Q})$ given in
such a scheme to the leading order by
\begin{equation}\label{inf11}
\Phi|_{extr}=\ln{L}\left\{\begin{array}{cc}1-2\gamma,  & \quad
0\le \gamma\le 1/3\\ \frac{(1-\gamma)^2}{2\gamma}, & \quad 1/3\le
\gamma<1
\end{array}\,\right.\,.
\end{equation}  The above result immediately implies
the values for the corresponding multifractality exponent
$\tilde{\tau_2}$ coinciding  everywhere in the high-temperature
phase $0\le \gamma<1$ with that exemplified by the standard REM,
cf. (\ref{REM}).

To complete the picture, one can perform the standard
de-Almeida-Thouless-like analysis\cite{AT} and verify that the
replica-symmetric solution Eq.(\ref{inf5}) does not show any local
instability at the point of the pre-freezing transition $\gamma\to
1/3-0$ (although it does become unstable for a certain value
$\gamma=\gamma_{inst}\in (1/3,1/2)$). Absence of local instability
is consistent with the mentioned observation that in spite of
$m\to 0$ when approaching the pre-freezing point the solution with
broken replica-symmetry (\ref{inf8}) remains formally different
from the replica-symmetric solution (\ref{inf5}).  Above the
pre-freezing temperature, that is for $\gamma<1/3$, the two
solutions share the same value of the functional $\Phi|_{extr}$,
and therefore produce exactly the same simple scaling for the
multifractality exponent $\tilde{\tau}_2$. In the pre-freezing
domain $1/3<\gamma<1$ the solution with broken replica symmetry
wins.

In conclusion, we have performed systematic analysis of the
multifractality exponents extracted from the averaged moments of
the Boltzmann-Gibbs measure generated by logarithmically
correlated random potentials. In particular, using
zero-dimensional and infinite-dimesnional versions of the model we
have identified a pattern of the replica symmetry breaking
responsible for the abrupt change ("pre-freezing") of those
exponents in the high-temperature phase. Implementing such a
pattern in explicit calculations for one- and two-dimensional
versions of the models remains a challenging open problem.

\subsubsection*{\sf Acknowledgements.} This work was supported by
Leverhulme Research Fellowship project "A single particle in
random energy landscapes". Major part of the research has been
completed during the author's participation in the programme on
Mathematics and Physics of Anderson Localisation at the Newton
Institute(Cambridge) whose support and hospitality is gratefully
acknowledged. The author is grateful to A. Mirlin for stimulating
initial discussions on the problem of annealed multifractality
exponents, and to C. Mudry, P. Le Doussal and M. Skvortsov for
encouraging interaction at various stages of the work.


\vskip 1.2cm

\begin{thebibliography}{99}
\bibitem{PV} Paladin G and Vulpiani A
{\it Phys. Rep.} {\bf  156} (1987) 147

\bibitem{ME} Evers F and Mirlin AD,  {\it Rev. Mod. Phys.} {\bf 80} (2008)
1355

\bibitem{FRL} Foster MS, Ryu S, and
Ludwig AWW (2009), e-preprint arXiv:0901.0284


\bibitem{MG} Monthus C and Garel T {\it Phys. Rev. E} {\bf 75}
(2007), Art. No. 051122

\bibitem{negf} Mandelbrot B {\it Physica A} {\bf 163} (1990),
306; Chabra AB and Sreenivasan KR {\it Phys. Rev. A} {\bf 43 }
(1990) 1114 ; Halsey TC, Honda K, and Duplantier B {\it J. Stat.
Phys.} {\bf 85} (1996) 681

\bibitem{CLD} Carpentier D and  Le Doussal P  {\it Phys. Rev. E} {\bf 63} (2001), 026110

\bibitem{FB1} Fyodorov YV and Bouchaud JP {\it J. Phys.A: Math.Theor} {\bf 41}
(2008) 324009 (25pp); Fyodorov YV and Sommers H-J {\it Nucl. Phys.
B [FS]} {\bf 764} (2007), 128

\bibitem{LFSG} Ludwig A, Fisher MPA, Shankar R, and Grinstein G
{\it Phys. Rev. B} {\bf 50} (1994), 7526

\bibitem{2d} Chamon C, Mudry C and Wen X-G {\it Phys. Rev. Lett. } {\bf 77} (1996)  4194;
Castillo H E, Chamon C C, Fradkin E, Goldbart P M, and Mudry C
{\it Phys. Rev. B } {\bf 56} (1997) 10668

\bibitem{Liouville} Kogan II, Mudry C, and Tsvelik AM {\it Phys.
Rev. Lett.} {\bf 77} (1996) 707

\bibitem{selfgrav}
Abdalla E and Tabar M R R  {\it Phys. Lett. B} {\bf 440} (1998)
339


\bibitem{FB2} Fyodorov YV and Bouchaud JP {\it J. Phys.A: Math.Theor} {\bf 41}
(2008) 372001 (12pp)

\bibitem{BMD} Muzy J-F, Delour J, and Bacry E {\it Eur. Phys. J. B} {\bf 17} (2000) 537
Bacry E, Delour J, and Muzy J-F {\it Phys. Rev. E} {\bf 64} (2001)
026103

\bibitem{Ostrov} Ostrovsky D {\it Lett. Math. Phys} {\bf 83}
(2008) 265

\bibitem{FLDR} Fyodorov YV, Le Doussal P, and Rosso A, under
preparation.

\bibitem{DS} Duplantier B and Sheffield S  (2008), e-preprint arXiv:0808.1560
[math.PR]


\bibitem{Vargas} Rhodes R and Vargas V (2008) e-preprint arXiv:0807.1036[math.PR]


\bibitem{REM} Derrida B {\it Phys. Rev. B} {\bf 24} (1981) 2613

\bibitem{GD89} Gardner E and Derrida B {\it J.Phys.A} {\bf 22} (1989) 1975

\bibitem{DT} Derrida B and Tolouse T {\it J Physique Lett} {\bf
46} (1985) L223

\bibitem{Derrida3} Derrida B {\it Physica D} {\bf 107} (1997) 186

\bibitem{BM}  Bouchaud J-P and M\'ezard M  {\it J. Phys. A: Math. Gen.} {\bf 30} (1997) 7997

\bibitem{Forrester} Forrester P J and Warnaar S O ,  Bull. Amer. Math. Soc. (N.S.) 45 (2008), no. 4, 489--534

\bibitem{AK} Klimovsky A (2009), "Parisi landscapes in high-dimensional Euclidean spaces", 
talk at the workhop "Mathematical Models from Physics and Biology", April 2009, Bonn, Germany

\bibitem{AT}  de Almeida JRL and Thouless DJ {\it J.Phys.A}
{\bf 11} (1978), 983
\end{thebibliography}
\end{document}